\documentclass[epj,english]{webofc}
\usepackage[varg]{txfonts}   
\usepackage{caption}
\usepackage[skip=0pt]{subcaption}
\wocname{EPJ Web of Conferences}
\woctitle{INPC 2013}

\begin{document}
\title{Mass dependence of nuclear short- range correlations in nuclei
and the EMC effect}
%
%

\author{Wim Cosyn\inst{1}\fnsep\thanks{\email{wim.cosyn@ugent.be}} \and
        Maarten Vanhalst\inst{1} \and
        Jan Ryckebusch\inst{1}
}

\institute{Department of Physics and Astronomy,\\
 Ghent University, Proeftuinstraat 86, B-9000 Gent, Belgium
          }

\abstract{%
  We sketch an approximate method to quantify the number of correlated pairs 
in any nucleus $A$. It is based on counting independent-particle model (IPM) 
nucleon-nucleon pairs in a relative $S$-state with no radial excitation.  We 
show that  IPM pairs with those quantum numbers are most prone to short-range 
correlations and are at the origin of the high-momentum tail of the nuclear 
momentum distributions.  Our method allows to compute the $a_2$ ratios extracted 
from inclusive electron scattering. Furthermore, our results reproduce the 
observed linear correlation between the number of correlated pairs and the 
magnitude
of the EMC effect.  We show that the width of the pair 
center-of-mass distribution in exclusive two-nucleon knockout yields 
information on the quantum numbers of the pairs.
}
\maketitle

One of the striking features of the nucleon-nucleon potential
is the appearance of a huge repulsive core in the potential for 
inter-nucleon distances of $\lesssim$ 1 fm.  This is a reflection of the 
finite size of the nucleon.  The hard core introduces short-range
correlations (SRC) in the nuclear quantum many-body system.  These cause
high-density 
and high-momentum fluctuations that are reflected in the fat tails of the 
one-body momentum distribution.  Recent theoretical and experimental efforts 
have identified proton-neutron pairs as the dominant contribution to SRC, an 
effect which can be attributed to  
the tensor part of the nuclear force (for recent reviews see 
\cite{Arrington:2011xs,Frankfurt:2008zv}). Somewhat surprisingly, a very nice 
linear correlation was also 
observed between the magnitude of the EMC (European Muon Collaboration) effect
for a nucleus $A$ and the ratios of  
inclusive electron scattering from nucleus $A$ relative to the deuteron (taken 
as a measure for the amount of correlated pairs in a nucleus) 
\cite{Weinstein:2010rt,Hen:2013oha}.  This might indicate that both phenomena 
share a common origin, being caused by high-density and/or high-virtuality 
fluctuations in the nucleus. Here, high-virtuality refers to high-momentum bound
nucleons, consequently far off the mass shell.

\begin{figure}[ht]
\centering
\includegraphics[width=7cm,clip]{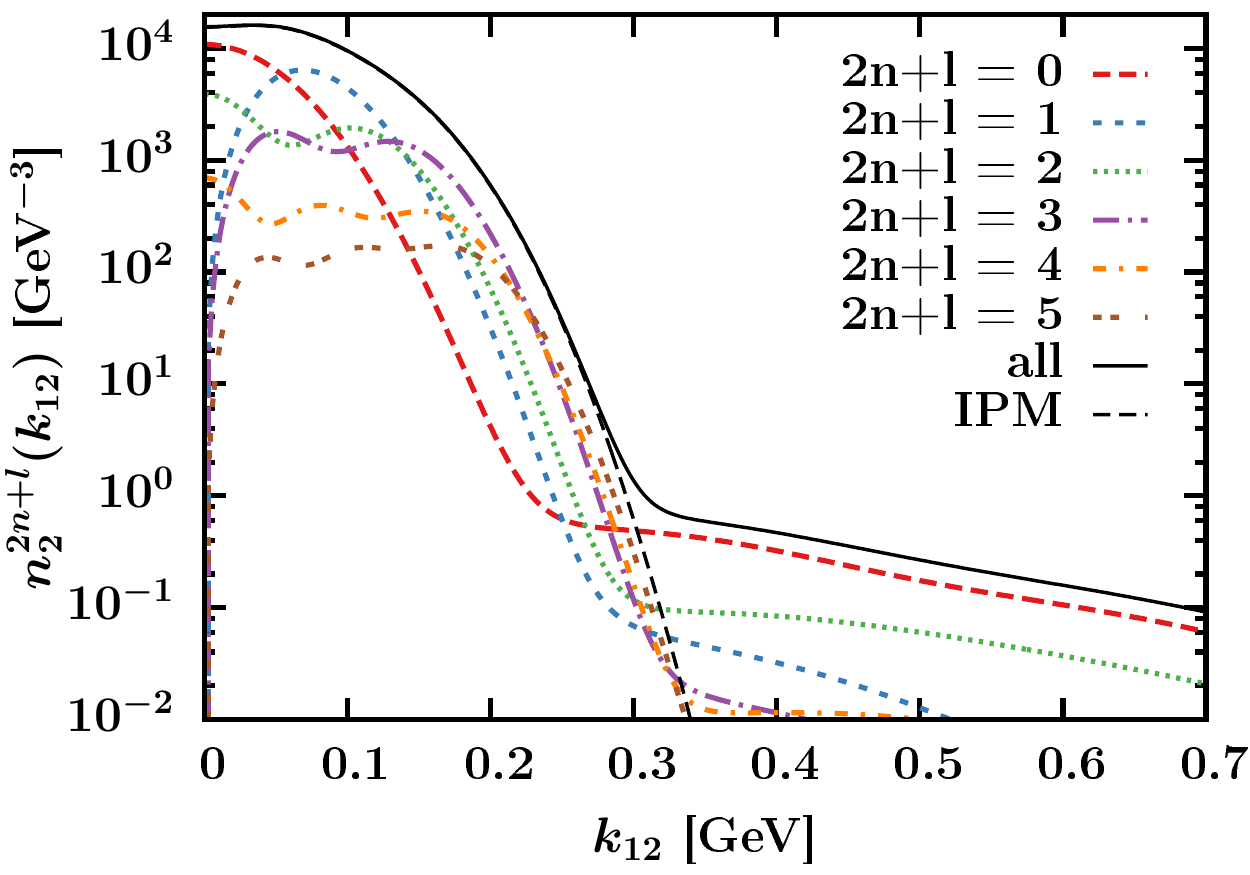}
\caption{(Color online)
 Momentum dependence of the relative two-body momentum 
distribution $n_2$ (solid black curve) in function of the relative
momentum $k_{12}$ of the two nucleons, as computed in the lowest-order cluster 
expansion for $^{56}$Fe.  The black dashed curve is the IPM prediction.  The 
other lines show the different contributions $n_2^{2n+l}$ to $n_2$ from the
correlation operators acting on 
two-nucleon states with various $nl$.}
\label{fig:Femom}       
\end{figure}

In Refs. \cite{Vanhalst:2011es,Vanhalst:2012ur,Vanhalst:2012zt}, we 
have introduced an approximate method to quantify the number of correlated 
pairs in any nucleus. 
We start from the well-known
approach where a 
symmetrical correlation operator $\widehat{\mathcal{G}}$ is used to transform a 
Slater 
determinant of independent particle model (IPM) single-particle wave functions 
$\Psi_A^\text{IPM}$ into a realistic wave function $ \Psi_A$
\cite{Pieper:1992gr}:
\begin{equation}
  \mid { \Psi_A}   \rangle =  \frac{1}
{ \sqrt{\langle \ \Psi  ^{IPM} _A \mid \widehat{\cal
G}^{\dagger} \widehat{\cal G} \mid \Psi  ^{IPM} _A \ \rangle}} \ 
\widehat
{ {\cal G}} \mid  \Psi  ^\text{IPM} _A \ \rangle \; .
\label{eq:realwf}
\end{equation}
The operator $\widehat{\cal{G}}$ is complicated but as far as the SRC
are concerned, it is dominated by the central, tensor and spin-isospin
correlations \cite{janssen00,ryckebusch97}.  We then exploit the
short-ranged nature of the dominant contributions in the following
way.  We start from a harmonic oscillator (HO) basis for the Slater
determinant $\Psi_A^\text{IPM}$.  This does not impose
restrictions on the choice of IPM model as any wave function can be
expanded in the HO basis.  Next, for the two-particle wave function,
we perform a transformation to relative and center-of-mass (c.o.m.)
coordinates using Talmi-Moshinsky brackets
\cite{moshinskyharmonic}. This allows us to identify the contributions
of all relative ($nl$) and c.o.m.  ($NL$) HO quantum numbers to the
two-particle wave functions \cite{Vanhalst:2011es,Vanhalst:2012ur}.
As the typical central and tensor correlation operators in
Eq.~(\ref{eq:realwf}) project on pairs with a substantial
close-proximity ($r\approx0$) probability, we identify the IPM
$(n=0,l=0)$ nucleon pairs as those parts of $\Psi_A^\text{IPM}$ prone
to SRC.  The validity of this assumption is nicely illustrated in
Fig. \ref{fig:Femom}, where we show the different $(n,l)$
contributions to the two-body momentum distribution in $^{56}$Fe as
computed in a lowest order cluster approximation of
Eq.~(\ref{eq:realwf}).  One clearly observes that the SRC generated
high-momentum tail in the momentum distribution is dominated by the
contribution from correlation operators acting on IPM pairs with
$(n=0,l=0)$ relative quantum numbers.  We wish to stress that the
high-momentum tails in the momentum distributions can have very
different quantum numbers than $(n=0,l=0)$ through the operation of the
correlation operator $\mathcal{G}$.  The most obvious example is the
dynamical generation of the deuteron D-wave $(l=2)$ as the tensor part
of $\widehat{\mathcal{G}}$ acts on the ``IPM'' S-wave $(l=0)$ component of the 
wave function.

\begin{figure}[ht]
\begin{subfigure}[b]{0.49\textwidth}
\centering
\includegraphics[width=\textwidth,clip]
{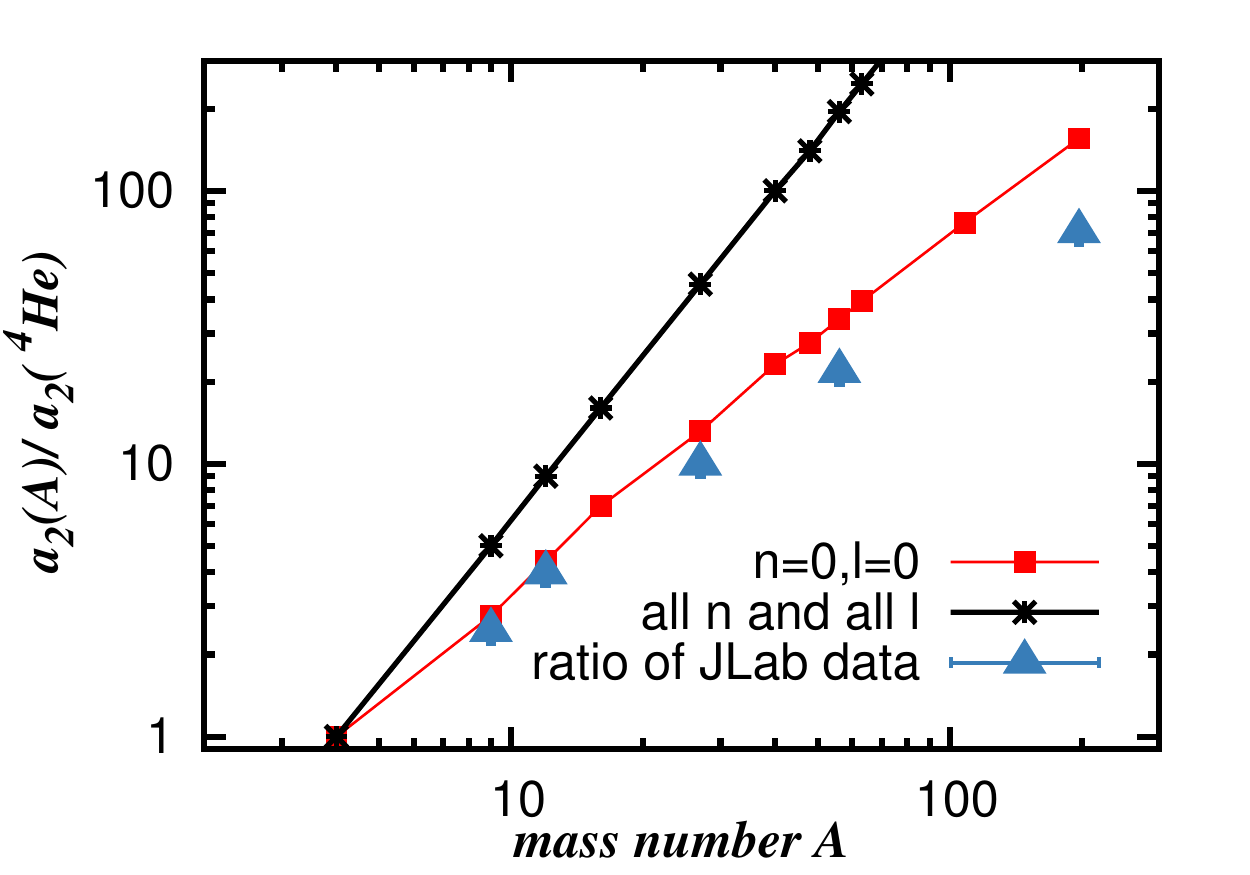}
\caption{}
\label{fig:a2ratio}
\end{subfigure}
\begin{subfigure}[b]{0.51\textwidth}
\centering
\includegraphics[angle=-90,width=\textwidth,clip]{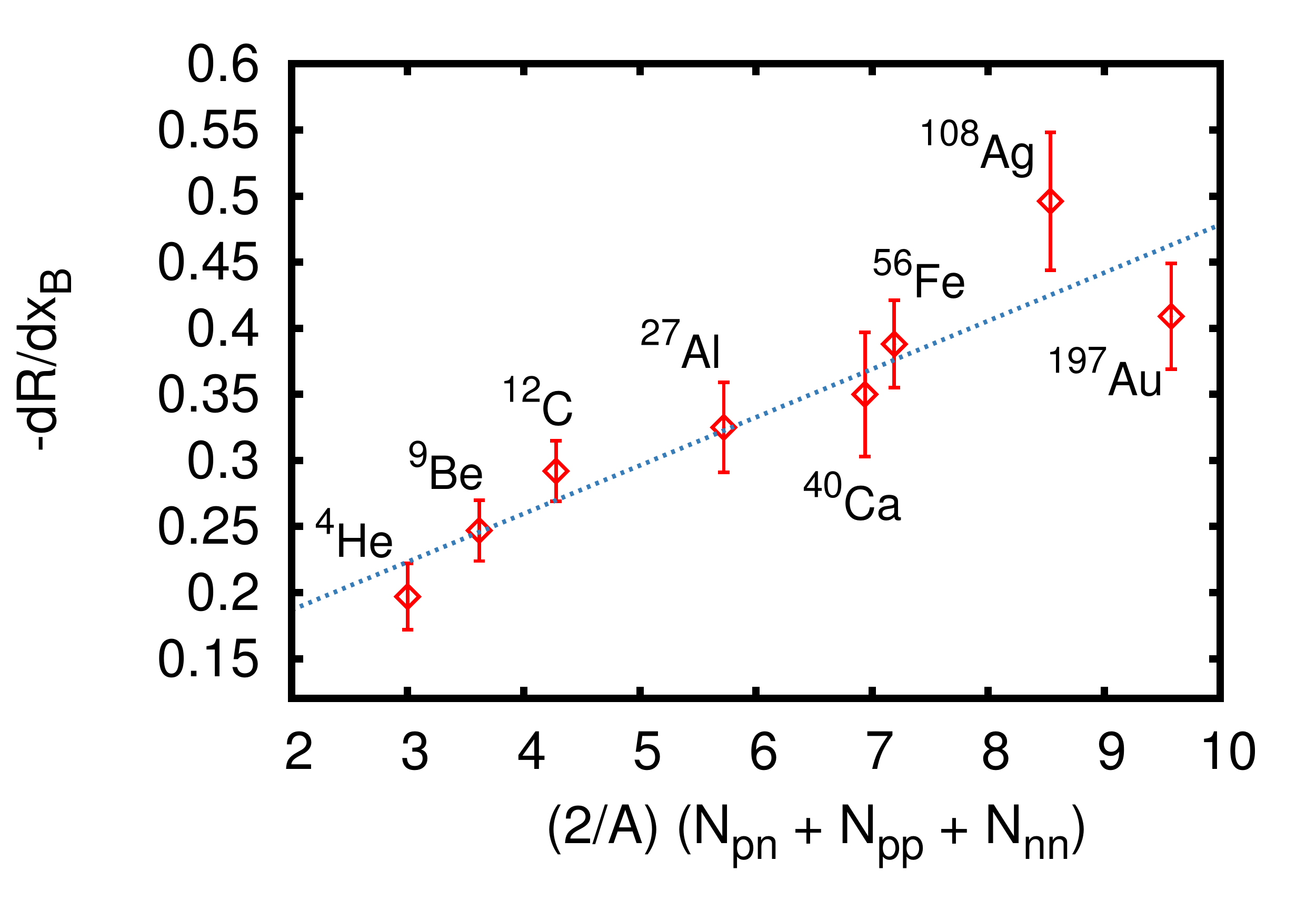}
\caption{}
\label{fig:EMC}
\end{subfigure}
\vspace{-1cm}
\caption{(Color online) (a) Mass dependence of the $a_2$ relative to $^{4}$He.
The 
black crosses are the predictions given that all relative  quantum numbers 
equally contribute.  For the red squares only $pn$ pairs in a triplet spin 
state and with $n=0,l=0$ are counted. The data are from Ref. 
\cite{Hen:2012fm}. The curves are guide to the eyes. (b) The magnitude of the
EMC effect versus
the computed number of correlated $NN$ pairs per nucleon. The EMC data are
from the analysis presented in Ref.~\cite{Hen:2012fm}.
}\label{fig:fig2combo}
\end{figure}

The mass dependence of nuclear SRC has experimentally been quantified in
inclusive electron scattering 
(for an overview of the 
world data see \cite{Hen:2012fm}).  The ratio 
of the inclusive cross
section at moderate $Q^2$  on a nucleus $A$ to the deuteron shows a 
scaling
plateau at values of Bjorken $x_B$ of $1.5 \lesssim x_B \lesssim 2$.  The value 
of the measured ratio rescaled by $\frac {2}{A}$ (a quantity denoted by 
$a_2(A)$) can be interpreted as a measure of the amount of correlated pairs in 
$A$ relative to the deuteron. 
We compare in Fig. \ref{fig:a2ratio} the experimental values of these $a_2$ in
superratios to ${^4}\text{He}$ with our calculations.  The kinematics in 
the inclusive experiments probe initial nucleon momenta in the range of 
300-500 MeV, where the tensor correlation function forms the dominant 
contribution to the high-momentum tail.  Thereby, in our calculations we 
compute the $a_2(A)$ by counting the close-proximity 
$(n=0,l=0)$ $np$ pairs in a spin triplet ($S=1$) state.
One can observe that the predictions are in acceptable agreement with
the slope and 
magnitude of the measured $A$-dependence 
data for light and medium nuclei, but tend to overestimate the measured values
at high $A$. No corrections 
accounting for the c.o.m. motion or final-state interactions of the pairs in 
the nucleus $A$ were applied to the calculations in Fig.~\ref{fig:a2ratio}, 
providing a possible explanation for this overestimation.   It is also 
obvious that the scaling
with $A$ of $a_2(A)$ is a lot softer than one would expect from the naive 
counting of all $np$ pairs, also depicted in Fig. \ref{fig:a2ratio}.

Given the experimentally established linear correlation between the size of the
EMC effect and the amount of correlated pairs in a nucleus, we want to check if
this correlation is also present when we compare our model calculations to
EMC data. In Fig. \ref{fig:EMC}, we display the magnitude of the EMC
effect - quantified by means of the slope $-dR/dx_B$ of the EMC ratio for
Bjorken $0.35 < x_B < 0.7$ \cite{Seely:2009gt} - versus our predictions for
the
probability for $NN$
SRC per nucleon relative to the deuteron. Here, contrary to the $a_2$ ratios
in 
Fig.~\ref{fig:a2ratio}, we compare to the combined number of $pp$, $np$ and $nn$ 
pairs.  This is motivated
by the partonic nature of the deep-inelastic scattering process (all
 pairs can contribute equally), whereas the inclusive
scattering is dominated by the tensor-correlated triplet-spin proton-neutron pairs. We
stress that the numbers
which one finds on the $x$ axis of Fig.~\ref{fig:a2ratio} are the results of parameter-free
calculations.    A nice linear relationship can be observed between the quantity
which we
propose as a per nucleon measure for the magnitude of the
SRC and the magnitude of the EMC effect. If the observed trends apply to the
whole nuclear mass range,
we could predict the size of the EMC effect for any nucleus $A$.


So far, in Fig. \ref{fig:fig2combo}, we have compared our model calculations to
quantities extracted from inclusive electron scattering experiments. To study
the precise nature of nuclear SRC, the exclusive $A(e,e'NN)$ reaction
provides more opportunities than inclusive measurements, but is of course
more demanding to measure.  Similarly to the scaling with the 
distorted
one-body momentum distribution for the quasi-elastic $A(e,e'N)$ cross section, it can be shown
that under kinematic conditions probing correlated nucleon-nucleon pairs, the 
$A(e,e'NN)$
cross section scales with the distorted c.o.m. momentum
distribution of close-proximity pairs.  When comparing 
the width of the computed c.o.m. momentum distribution restricted to the
$(n=0,l=0)$ pairs, with the width of the distribution for all 
pairs, one observes
a significant difference between the two.  The numbers for the widths
of the c.o.m. distributions for several relative $l$ quantum numbers in 
$^{12}$C are listed in Table~\ref{tab:widths}.  The width of the c.o.m. 
distribution for close-proximity $l=0$ pairs is 14 MeV wider than the one for 
the distribution which does not impose restrictions on the $l$.

\begin{table}
\centering
\caption{The width of the $^{12}$C c.m. distribution for
    pp pairs with relative orbital momentum $l$.}
\label{tab:widths}       
\begin{tabular}{ccccc}
\hline
& $l=0$ & $l=1$ & $l=2$ & all $l$   \\\hline
$\sigma~[MeV]$ & $154$ & $135$ & $121$ & $140$ \\\hline
\end{tabular}
\end{table}

To conclude, we have shown that the amount of correlated pairs in any nucleus
can be approximately quantified by counting the number of relative $(n=0,l=0)$
pairs in the IPM wave function.  Our calculations for the $a_2$ ratios are
consistent with the data for light and medium nuclei, but tend to overestimate
these at high $A$. 
We observe a
nice correlation between the computed number of correlated pairs and the 
measured magnitude of the EMC effect.  We have shown that the width of the 
c.o.m. momentum distribution differs significantly between all possible and 
correlated nucleon-nucleon pairs, the latter being 14 MeV wider in $^{12}$C.  
This is a quantity that can be accessed in exclusive $A(e,e'NN)$ reactions 
at moderate $Q^2$.

\begin{acknowledgement}
This work is supported by the Research
Foundation Flanders (FWO-Flanders) and by the Interuniversity
Attraction Poles Programme initiated by the Belgian Science Policy
Office. The computational resources (Stevin Supercomputer
Infrastructure) and services used in this work were provided by Ghent
University, the Hercules Foundation and the Flemish Government –
department EWI. 
\end{acknowledgement}

%
\bibliography{INPC13.bib}

\end{document}